\documentclass[preprint,aps,superscriptaddress,preprintnumbers,amsmath,amssymb,prl]{revtex4-1}
\usepackage{amssymb}
\usepackage{amsmath}
\usepackage{graphicx}
\usepackage{multirow}
\usepackage{comment}
\usepackage{longtable}
\usepackage{color}
\usepackage{xfrac}
\graphicspath{{./fig}}

\begin{document}
\newcommand{\tny}[1]{\mbox{\tiny $#1$}}
\newcommand{\eqn}[1]{\mbox{Eq.\hspace{1pt}(\ref{#1})}}
\newcommand{\eqs}[2]{\mbox{Eq.\hspace{1pt}(\ref{#1}--\ref{#2})}}
\newcommand{\eqsu}[2]{\mbox{Eqs.\hspace{1pt}(\ref{#1},\ref{#2})}}
\newcommand{\eqtn}[2]{\begin{equation} \label{#1} #2 \end{equation}}
\newcommand{\func}[1]{#1 \left[ \rho \right] }
\newcommand{\mfunc}[2]{#1_{#2} \left[ \rho \right] }
\newcommand{\mmfunc}[3]{#1_{#2} \left[ #3 \right] }
\newcommand{\mmmfunc}[4]{#1_{#2}^{#3} \left[ #4 \right] }
\newcommand{\pot}[1]{v_{\rm #1}}
\newcommand{\spot}[2]{v_{\rm #1}^{#2}}
\newcommand{\code}[1]{\texttt{#1}}

\def\brpppp{{\mathbf{r}^{\prime\prime\prime\prime}}}
\def\brppp{{\mathbf{r}^{\prime\prime\prime}}}
\def\brpp{{\mathbf{r}^{\prime\prime}}}
\def\brp{{\mathbf{r}^{\prime}}}
\def\bzp{{\mathbf{z}^{\prime}}}
\def\bxp{{\mathbf{x}^{\prime}}}
\def\tp{{{t}^{\prime}}}
\def\tpp{{{t}^{\prime\prime}}}
\def\tppp{{{t}^{\prime\prime\prime}}}

\def\tbr{{\tilde{\mathbf{r}}}}
\def\bk{{\mathbf{k}}}
\def\br{{\mathbf{r}}}
\def\bz{{\mathbf{z}}}
\def\bx{{\mathbf{x}}}
\def\bR{{\mathbf{R}}}
\def\bM{{\mathbf{M}}}
\def\bP{{\mathbf{P}}}
\def\bT{{\mathbf{T}}}
\def\bK{{\mathbf{K}}}
\def\bA{{\mathbf{A}}}
\def\bB{{\mathbf{B}}}
\def\bX{{\mathbf{X}}}
\def\bY{{\mathbf{Y}}}
\def\bP{{\mathbf{P}}}
\def\bI{{\mathbf{I}}}
\def\d{{\mathrm{d}}}
\def\rhor{{\rho({\bf r})}}
\def\rhorp{{\rho({\bf r}^{\prime})}}
\def\rhoi{{\rho_I}}
\def\rhoii{{\rho_{II}}}
\def\rhoj{{\rho_J}}
\def\rhoir{{\rho_I({\bf r})}}
\def\rhoiir{{\rho_{II}({\bf r})}}
\def\rhojr{{\rho_J({\bf r})}}
\def\rhoirp{{\rho_I({\bf r}^{\prime})}}
\def\rhojrp{{\rho_J({\bf r}^{\prime})}}
\def\sumi{{\sum_I^{N_S}}}
\def\sumj{{\sum_J^{N_S}}}
\def\im{{\operatorname{Im}}}

\def\etal{{\it et al.}}
\def\vdw{{van der Waals}}
\def\qe{{\sc Quantum ESPRESSO}}
\def\se{{Schr\"{o}dinger equation}}
\def\ses{{Schr\"{o}dinger equations}}
\def\bnabla{{\boldsymbol{\nabla}}}
\def\bchi{{\boldsymbol\chi}}
\def\bLambda{{\boldsymbol\Lambda}}
\def\bDelta{{\boldsymbol\Delta}}

\def\tfvw{{TF+$\frac{1}{5}$vW}}

\title{{\fontsize{16}{18}\selectfont Orbital-Free DFT Correctly Models Quantum Dots When Asymptotics, Nonlocality and Nonhomogeneity Are Accounted For}}

\author{Wenhui Mi$^\dagger$}
\affiliation{Department of Physics, Rutgers University, Newark, NJ 07102, USA}
\altaffiliation{Also affiliated with the Department of Chemistry, Rutgers University, Newark, NJ 07102, USA}
\author{Michele Pavanello$^\dagger$}
\email{m.pavanello@rutgers.edu}
\affiliation{Department of Physics, Rutgers University, Newark, NJ 07102, USA}
\altaffiliation{Also affiliated with the Department of Chemistry, Rutgers University, Newark, NJ 07102, USA}

\begin{abstract}
{
Million-atom quantum simulations are in principle feasible with Orbital-Free Density Functional Theory (OF-DFT) because the algorithms only require simple functional minimizations with respect to the electron density function. In this context, OF-DFT has been useful for simulations of warm dense matter, plasma, cold metals and alloys. Unfortunately, systems as important as quantum dots and clusters (having highly inhomogeneous electron densities) still fall outside OF-DFT's range of applicability. In this work, we address this century old problem by devising and implementing an accurate, transferable and universal family of nonlocal Kinetic Energy density functionals that feature correct asymptotics and can handle highly inhomogenous electron densities. For the first time to date, we show that OF-DFT achieves close to chemical accuracy for the electronic energy and reproduces the electron density to about 5\% of the benchmark for semiconductor quantum dots and metal clusters. Therefore, this work demonstrates that OF-DFT is no longer limited to simulations of systems with nearly homogeneous electron density but it can venture into simulations of clusters and quantum dots with applicability to rational design of novel materials.}
\end{abstract}
\maketitle
\newpage

Metal clusters and quantum dots constitute an important class of systems of pivotal importance for materials design particularly in photovoltaics \cite{Lan_2014}, catalysis \cite{Tyo_2015}, and even quantum computing \cite{Huo_2013}. Although these fields are already strongly shaped by computer-aided design, the high computational cost of available quantum-mechanical methods such as Kohn--Sham density functional theory (KS-DFT) \cite{Maitra_2016,burk2012} is hampering futher progress. In this playing field, what is really needed is a breakthrough in techniques alternative to the current standard, and among them \cite{Giese_2017,VandeVondele_2012,krish2015a,yang1990,jaco2014,Fedorov_2017,goed1999} Orbital-Free Density Functional Theory (OF-DFT) is a promising candidate. 

OF-DFT is a promising and intriguing alternative because approximate density functionals for the kinetic energy entirely replace the need to solve for a Schr\"{o}dinger equation. This completely bypasses its inherent complexity. Particularly, OF-DFT algorithms are promising because they involve a computational scaling of at most $\mathcal{O}(N\ln N)$, where $N$ is a measure of the system size, and a memory requirement of only $\mathcal{O}(N)$ \cite{wesolowski2013recent,karasiev2012issues,witt2018orbital}.

Unfortunately, even though OF-DFT has already proven to be successful for simulations of million-atom systems involving crystalline and liquid metals and alloys \cite{witt2018orbital,shao2018large,hung2009accurate,PhysRevLett.92.085501}, as well as plasmas and warm dense matter \cite{PhysRevLett.111.175002,PhysRevLett.121.145001,PhysRevLett.113.155006}, its applicability has been severely limited by the accuracy of the available Kinetic Energy density functionals (KEDFs). For example, finite systems such as metal clusters and quantum dots have been outside the range of applicability of OF-DFT.

In this work, we achieve a breakthrouth by carefully balancing three important aspects defining the KEDFs: asymptotics of the corresponding potential, intrinsic nonlocality, and ability to handle nonhomogeneous systems. Thus, already at conception, we make sure that the functionals are nonlocal, that their asymptotics matches the known exact behavior, and that their nonlocal kernels adapt to such large density inhomogeneities as the ones occurring at the interface of nonperiodic systems with the vacuum. For finite systems, such as clusters, the latter is perhaps the most important aspect, which we tackle head on.

In the following, we cast our KEDF development in the current stat-of-the-art and derive the main theoretical and algorithmmical developments. Afterwards, we benchmark the resulting density functionals by carrying out OF-DFT simulations on 4 metal clusters and 4 semiconductor quantum dots realized in one hundred possible geometries for each, spanning energy windows of up to several tens of eV.

Over the past two decades there have been tremendous advances in OF-DFT development. Various KEDFs have been proposed \cite{constantin2018semilocal,luo2018simple,Karasiev_2015,Trickey_2009,Karasiev_2009,Karasiev_2013,wang1998,wang1999,perr1994,lari2014,Smiga_2017,chai2007,xia2012,xia2012b,huan2010,ho2008,zhou2005,carl2003}. The majority of these functionals are appropriate for main-group metallic bulk systems, and some show potential for modeling bulk semiconductors \cite{huan2010,mi2018nonlocal}.  
Although semilocal KEDFs \cite{Karasiev_2015,Trickey_2009,Karasiev_2009} have seen a recent resurgence \cite{luo2018simple,constantin2018semilocal}, nonlocal KEDFs (such as WGC \cite{wang1998,wang1999}, HC \cite{huan2010}, WT \cite{wang1992},  MGP \cite{mi2018nonlocal}, and others \cite{Pearson_1993,smar1994,perr1994}) have historically delivered better results, particularly for bulk solids. An inspiring study by Chan, Cohen and Handy found semilocal KEDFs to be theoretically appropriate for applications to clusters \cite{chan2001}. This was followed by several works on metallic clusters \cite{aguado2001orbital,aguado2001melting,aguado1999orbital} which employed simple combinations of Thomas--Fermi (TF) and fractions of von Weizs\"acker (vW) KEDFs  (e.g., $\frac{1}{9}$ or $\frac{1}{5}$). Unfortunately, the conclusions of those studies were mixed. Thus, in this work we depart from semilocal KEDFs and adopt fully nonlocal ones exploiting the typical nonlocal KEDF ansatz,
\begin{equation}
\label{tkin}
\func{T}=\mfunc{T}{\rm TF}+\mfunc{T}{\rm vW}+\mfunc{T}{\rm NL},
\end{equation}   
where $T_{\rm TF}$ is the Thomas--Fermi kinetic energy \cite{fermi1927,thom1927}, $\mfunc{T}{\rm vW}$ is the von Weizs\"acker functional \cite{weiz1935}, and $\mfunc{T}{\rm NL}$ is the remaining nonlocal contribution. The general form of $T_{\rm NL}$ is
\begin{equation}
\label{tnadd1}
\mfunc{T}{\rm NL}=\int  \rho^\alpha\left( \br \right) 
\omega_{\rm NL}\left[ \rho \right]\left( \br, \brp \right) \rho^\beta \left( \brp \right)\d\br\d\brp.
\end{equation}
Where, $\omega_{\rm NL}\left[ \rho \right]\left( \br, \brp \right)$ is the kernel of the nonlocal KEDF, and $\alpha$ and $\beta$ are positive scalars.

Let us first describe details of the MGP functional and then outline the new developments that extend its applicability to finite systems. Following the lead of Kohn and Sham \cite{kohn1965}, the starting point of our derivations is the linear response of the Free Electron Gas. This starting point is common among nonlocal functionals \cite{wang2000}. After a procedure of functional integration, this yields the WT functional \cite{wang1992} in the zero-th order and the MGP functional at higher orders \cite{mi2018nonlocal}. MGP's kernel correctly behaves in the low $q$ limit by construction, as we impose the so-called ``kinetic electron'' ({\it vide infra}) and therefore it can potentially approach systems beyond bulk metals. In our previous work \cite{mi2018nonlocal}, we found that MGP reproduces with remarkable accuracy the energetic properties and electron densities of silicon and several III--V semiconductors provided that two free parameters are adjusted. 

MGP's nonlocal potential is given by
\begin{align}
\label{eq:potential}
v_{\rm NL}(\br) = \rho(\br)^{-\frac{1}{6}}\hat{F}^{-1}\left[ \hat{F}\left[ \rho^{\frac{5}{6}} \right](q)~ \omega_{MGP}(q) \right](\br),
\end{align}
where the reciprocal space variable, $q$, corresponds to $|\br-\brp|$, and $\hat F[ \cdot ]$ stands for Fourier transform. Thus, the inherent approximation in \eqn{eq:potential} is that the kernel only depends on the magnitude of $|\br-\brp|$.  The kernel introduced in \eqn{eq:potential}, takes the following form 
 \eqtn{eq:kernel}{\omega_{MGP}(q)=\omega_{WT}(q)+\omega_{Hyper}(q)+\omega_{e}(q).}

The first term, $\omega_{WT}$, is the kernel of the WT functional. The second term, $\omega_{Hyper}$, originates from functional integration which, for historical reasons \cite{Hessler_1999}, we also call hypercorrelation \cite{mi2018nonlocal}. The functional integration step is employed to transform the kernel borrowed from the response function of the free electron gas into a kernel that can be used in the computation of the KEDF potential. In practice, the integration is carried out numerically after an integration by parts (see additional details in Ref.\ \citenum{mi2018nonlocal}). The third term, $\omega_{e}$, is the contribution encoding the kinetic electron and needs to be approximated. The kinetic electron arises from the long-range behavior of the exchange potential \cite{mi2018nonlocal}. An exact condition for the exchange potential in finite systems is that the negative of its source (i.e., $\frac{1}{4\pi}\nabla^2 v_{xc}(r)$) integrates to unity. As exchange potential and KEDF potential are connected {\it via} the Euler equation of DFT, the source of the KEDF (i.e., the kinetic electron) shall integrate to the opposite of the exchange hole. 

In contrast to Ref. \citenum{mi2018nonlocal}, here we propose a universal form for $\omega_{e}$, containing no adjustable parameters. Namely,
\eqtn{eq:KE}{\omega_{e}(q)=\frac{4\pi a}{q^{2}}\mathrm{erf}^{2}(q)\exp(-aq^{2}),}
where $a$ is a parameter that we relate to the number of electrons, $a=A/N_{e}^{2/3}$, with $A=0.2$ and $N_{e}$ is the total number of electrons.

Thus, if the kernel includes $\omega_{WT}(q)$, $\omega_{WT}(q)+\omega_{Hyper}(q)$ or $\omega_{WT}(q)+\omega_{Hyper}(q)+\omega_{e}(q)$ in Eq.(\ref{eq:kernel}), then the corresponding KEDF is called WT, MGP0, or MGP.  These kernels are only dependent on the average electron density, $\rho_{0}$, through $\eta=q/(2k_{F})$, where $k_{F}=(3\pi^{2} \rho_{0})^{1/3}$ is the Fermi wave vector associated with the average density. This approximation is too strong and is the source reason for needing adjustable parameters in these functionals. 
Thus, the kernel should be made dependent on the total electron density, $\rho(\br)$, instead of $\rho_{0}$. In principle, this would benefit and improve the description of systems where the distribution of electron density strongly deviates from uniformity.  

\begin{figure}[htp]
\begin{center}
\includegraphics[width=0.75\textwidth]{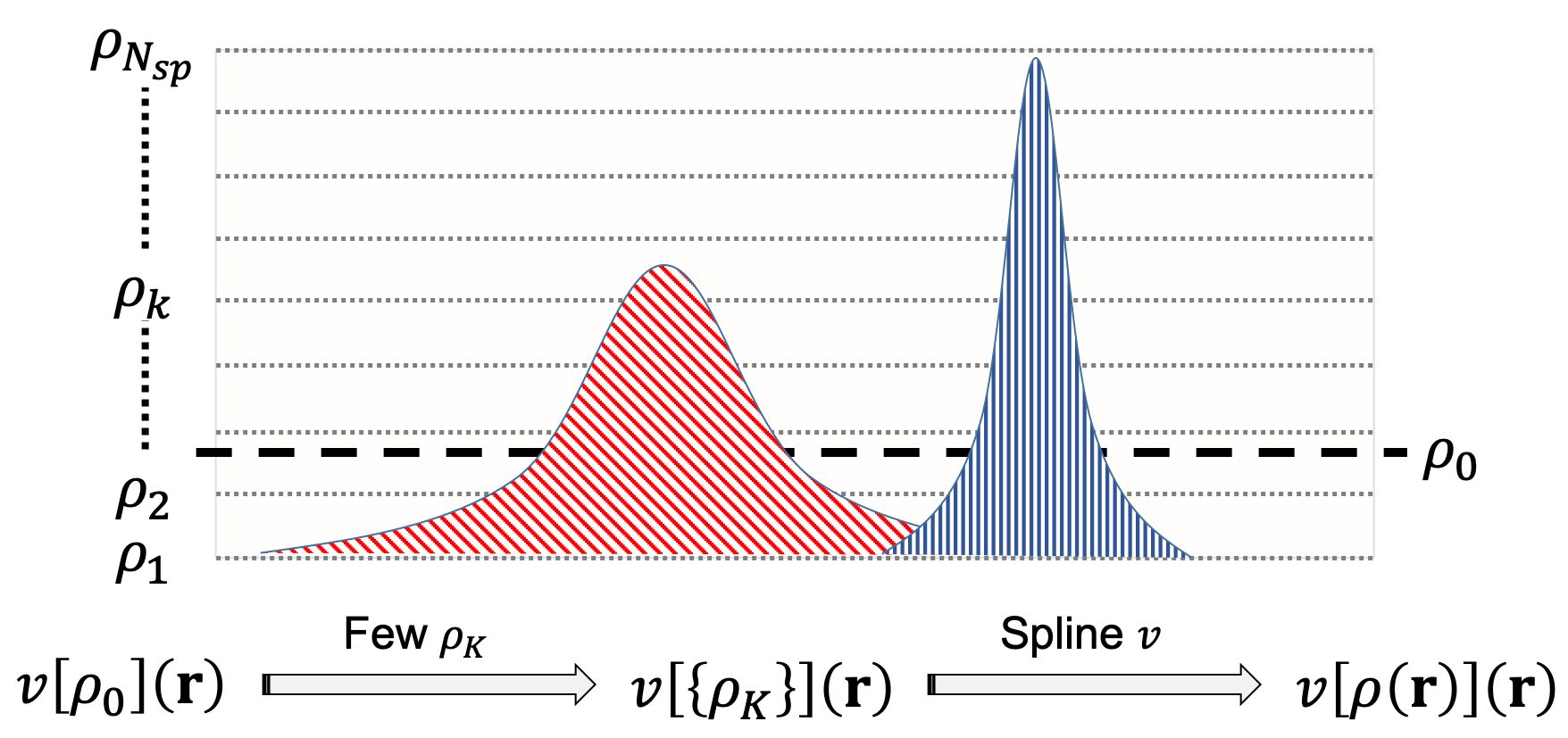}
\end{center}
\caption{\label{LDA} Two very different electron density profiles yield the same average density, $\rho_0$. We propose a generalization of nonlocal KEDFs with $\rho_0$-dependent kernel to become dependent on the full $\rho(\br)$ by evaluating the potential with a kernel built with $\rho_0=\rho_k$, $v[\rho_k](\br)$, for a set of $\big\{\rho_k\big\}_{k\in 1\ldots N_{sp}}$. Afterwards, extend to all possible values of $\rho(\br)$ with splines.}
\end{figure}

Such a strong approximation is shared among most nonlocal functionals. In this work we propose a method that tackles this limitation and in Figure \ref{LDA} we hint at the proposed workaround.
Inspired by the success of the local density approximation (LDA) \cite{Ceperley_1980,Perdew_1992}, we assume that the potential at a point $\br$ in space can be approximated by the one of a nonlocal functional evaluated with a kernel $\omega[\rho_0=\rho(\br)](q)$. This is the same principle as LDA applied to nonlocal kernels rather than to the energy densities, as customarily done. Unfortunately, implementing this principle directly would imply a super-quadratic computational cost ($N$ kernel evaluations each costing $N\ln N$). Fortunately, the computational scaling can be brought back to $\mathcal{O}\big(N\ln N\big)$ by employing spline techniques. Figure \ref{LDA} provides a visual depiction of the proposed spline method which we explain in detail in the following.

A series of $\rho_0$ values is considered, $\{\rho_{k}\}$, obtained by dividing the range between 0 and $\rho_{max}=\text{max}[\rho(\br)]$, in equally spaced segments
and choosing the total number of $k$ points to be $N_{sp}=40$. For each $\rho_{k}$, there is a corresponding kernel, $\omega_{\rm NL}(q,\eta(\rho_{k}))$, that can be tabulated and recovered ahead of computing the potential. Thus, a series of nonlocal potentials is obtained based on \eqn{eq:potential}, $\big\{v_{{\rm NL}}[\rho_k](\br)\big\}_{k\in 1\ldots N_{sp}}$, and the LDA procedure can be exploited invoking splines,
\begin{align}
v_{{\rm NL}}[\rho(\br)](\br) \xleftarrow{~\rm spline~} \bigg\{v_{{\rm NL}}[\rho_1](\br),\ldots, v_{{\rm NL}}[\rho_k](\br),\ldots,v_{{\rm NL}}[\rho_{N_{sp}}](\br)\bigg\}.
\end{align}
This is a scheme for constructing LDA versions of MGP (LMGP), WT (LWT) and MGP0 (LMGP0) functionals from the corresponding kernels.

Finally, The nonlocal contribution to the total kinetic energy is recovered by a second functional integration  
\begin{align}
\label{energy}
T_{\rm NL}[\rho] =\int d\br\int dt~\rho(\br)v_{{\rm NL}}[t\rho](\br).
\end{align}

Two other prescriptions for generating density-dependent kernels exist. WGC \cite{wang1999} exploits a Taylor expansion for the kernel around a reasonable value near $\rho_{0}$. Unfortunately, this can result in numerical instabilities when the electron density distribution differs much from the one of the free electron gas. Another example is  the kernel of the HC functional \cite{huan2010}. It is obtained by solving a differential equation when the electron density is updated. To ameliorate the computational cost, Huang and Carter offer an implementation of HC also featuring a spline technique in the spirit of Soler and coworkers \cite{roma2009}. Despite this, the computational cost of HC compared to WGC is still orders of magnitude larger. In terms of performance, the WGC functional can reproduce well KS-DFT results for main group bulk metals; HC functional can achieve significant improvement over previous functionals for semiconductors, showing promise for simulating finite systems when its two free parameters ($\lambda$ and $\beta$) are appropriately adjusted \cite{xia2012}. 

A major advantage of the LWT and LMGP family of functionals compared to WGC and HC is the fact that they are \emph{universal functionals} with no adjustable parameters. One issue with universal functionals is that they may not be transferable. I.e., they may work well for a certain system, but less well for others. In the following, we craft strict and conservative benchmarks for the proposed functionals that undeniably show their superiority compared to the current state-of-the-art and their transferability among an array of cluster sizes and types.

To assess the accuracy of the new family of KEDFs, we choose random clusters generated by CALYPSO \cite{CALYPSO_CPC,CALYPSO_PRB,CALYPSO_cluster}. 
Standard KS-DFT calculations provide benchmark values for the total energy and electron density (KS-DFT employs the exact KEDF) are carried out with Quantum ESPRESSO (QE) \cite{qe}. The OF-DFT simulations are performed with a modified version of ATLAS \cite{ATLAS,shao2016n} and PROFESS~3.0 \cite{PROFESS3}. To avoid dealing with nonlocal pseudopotentials, the optimal effective local pseudopotentials (OEPP) \cite{OEPP}  are employed for both OF-DFT and KS-DFT calculations.  LDA exchange-correlation energy functional by Perdew and Zunger \cite{Perdew1981} is employed in all calculations. Additional technical and computational details are available in the Supplementary Materials.

\begin{figure}[htp]
\begin{center}
\includegraphics[width=1.0\textwidth]{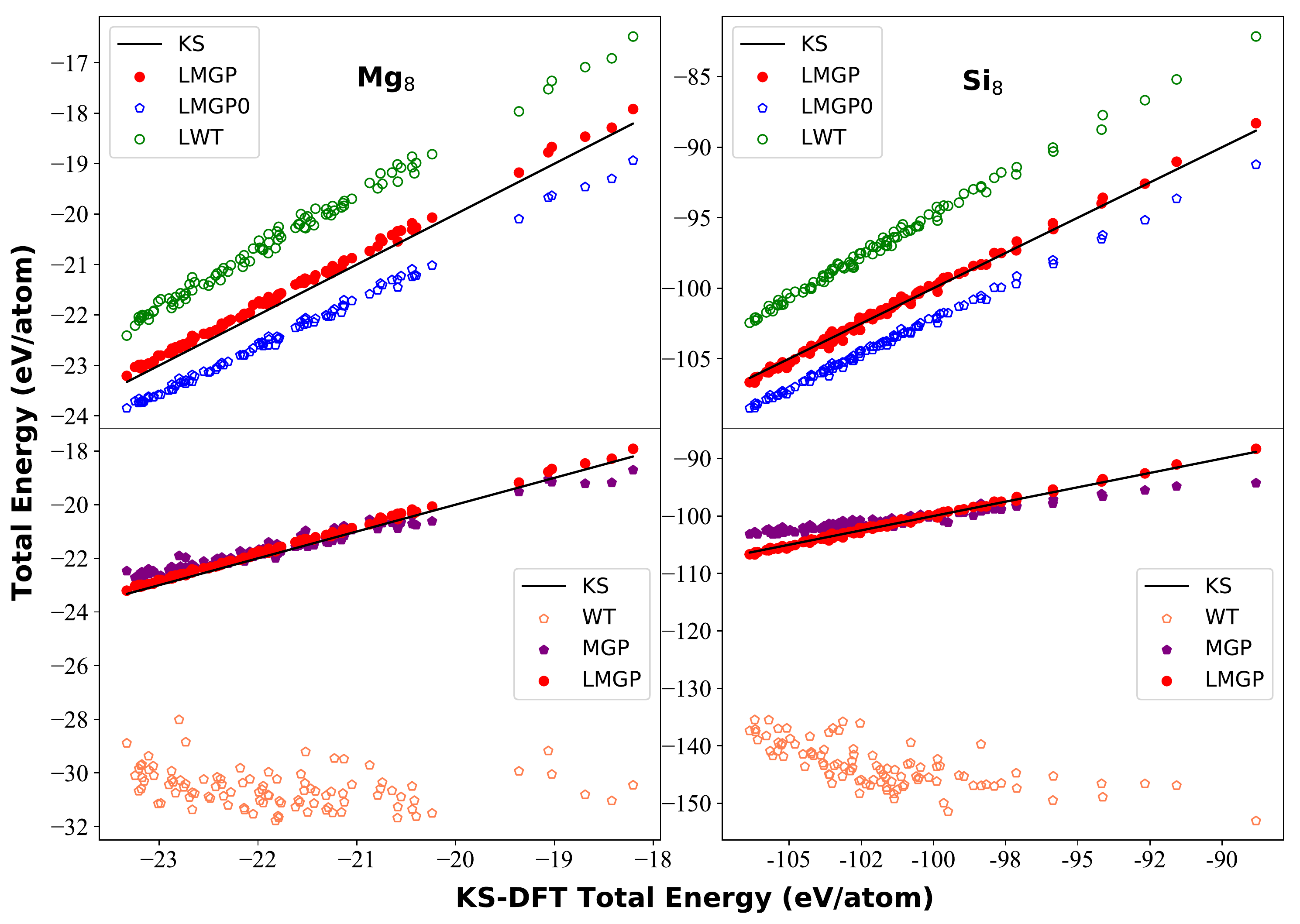}
\end{center}
\caption{\label{fig:Step} Total energies computed with WT, MGP, LWT, LMGP0, and LMGP KEDFs compared to reference KS-DFT values (on the diagonal) for 100 random structures of Mg$_8$ (left) and Si$_8$ (right) clusters.}
\end{figure}

With KS-DFT quantities as reference, we initially select two types of clusters: Mg$_8$ and Si$_8$. For each, we compute the total energy of 100 random structures and collect the computed values in Figure \ref{fig:Step}. The figure shows a progressive improvement when adopting the functionals WT $\to$ MGP $\to$ LMGP. In particular, we note that the consistent bias of MGP (slope differing from KS-DFT) is eliminated by the LDA procedure in LMGP. As shown in the lower insets of Figure \ref{fig:Step}, MGP improves dramatically total energies in comparison with WT. Furthermore, the three new parameterless functionals (LWT, LMGP0, and LMGP) are found to outperform their $\rho_0$-dependent kernel counterparts. We should note that while LWT and LMGP/LMGP0 functionals are universal (no adjustable parameters), MGP results are obtained by adjusting the parameter $a$ independently for Mg$_8$ and Si$_8$ clusters ($a=0.35$ and $0.5$, respectively). Strikingly, LMGP values are found to be essentially on-top of the KS-DFT results, providing us with an indication that the LDA procedure implemented by splines is stable and accurate for these systems.

\begin{figure*}[htp]
\begin{center}
\includegraphics[width=1.0\textwidth]{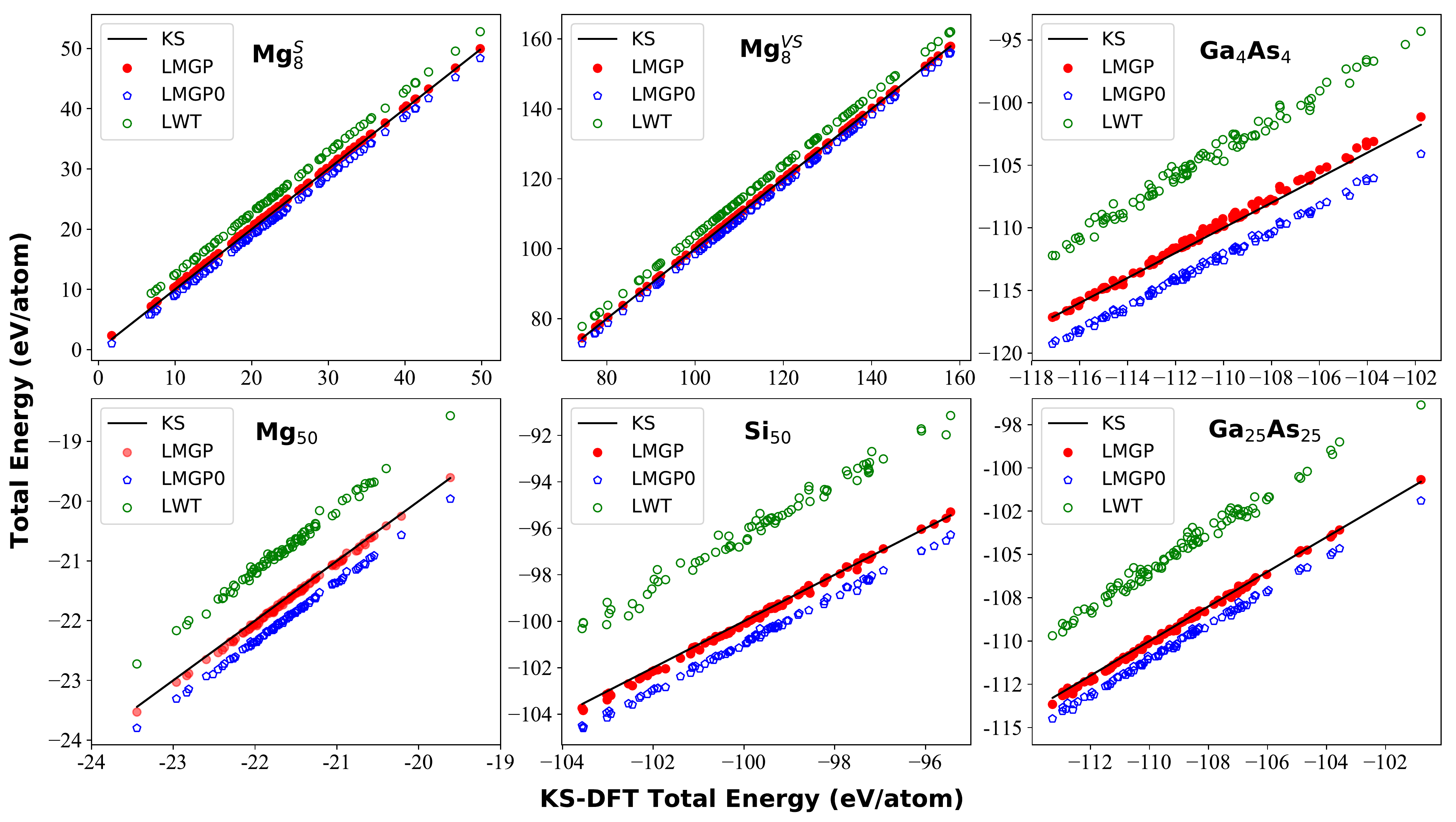}
\end{center}
\caption{ \label{fig:transferability}The total energies obtained by OF-DFT employing LWT, LMGP0, and LMGP KEDFs in comparison with the reference KS-DFT results for six different  cluster systems, Mg$_{50}$, Si$_{50}$, Ga$_4$As$_4$, Ga$_{25}$As$_{25}$, and two Mg$_{8}$ systems with shorter average bond distances: Mg$_8^\text{S}$ (i.e., strained) and Mg$_8^\text{VS}$ (i.e., very strained), respectively. For each system we generate 100 random structures.}
\end{figure*}

To confirm the transferability of our new functionals, we select six additional cluster systems: Mg$_{50}$, Si$_{50}$, Ga$_4$As$_4$, Ga$_{25}$As$_{25}$, and two more Mg$_{8}$ (i.e., Mg$_8^S$ and Mg$_8^{VS}$) featuring shorter average interatomic distances. The new set of systems provides us with an array of metallic to semiconducting quantum dot-like clusters. As shown in Figure \ref{fig:transferability}, the performance of our new functionals is consistent for all systems reproducing total energies across a wide window of energy spanning several eV per atom. In terms of absolute values of total energies, LWT and LMGP0  results are higher and lower than KS-DFT results, respectively. These results are source of considerable excitement -- not only the LMGP energy values correlate almost perfectly with the KS-DFT benchmark, but more importantly the LDA procedure (which is numerical in nature) is found to be stable for all systems considered. LMGP converges for more than 90\% of the structures in all systems with an average convergence rate of over 95\%. 

To quantify the performance of our new functionals, Table \ref{tab:Energy} shows the mean unsigned error (MUE) of total OF-DFT energies compared to KS-DFT for the 100 random structures of each system computed with LWT, LMGP, and LMGP0, as well as WT, and \tfvw.  

An option is to also compare against the Huang-Carter \cite{huan2010} (HC) and the Wang-Govind-Carter (WGC) functionals. However, even though HC is considered to be the most accurate KEDF presently available, it is also known for drawbacks that make it unsuitable for applications to finite systems \cite{xia2012}. Xia and Carter \cite{xia2012} found that especially for molecules and solid surfaces, the computational cost of HC is hundreds of times higher than the WGC functional. Additionally, 
despite our best efforts, HC (with $\lambda > 0$, an appropriate value for systems with gap, such as clusters) as well as WGC (second-order Taylor expansion of the kernel) did not converge for more than 10 of the 100 cluster structures considered for all systems. Thus, in this work, we compare against the other functionals. 

\begin{table}[!h]
\caption{\label{tab:Energy}  Mean unsigned error (MUE) for the total energy compared to KS-DFT in eV/atom. Superscripts $S$ and $VS$ stand for ``strained'' and ``very strained'', respectively.}
{\fontsize{10}{10}\selectfont 
\begin{tabular}{lccccc}
\hline \hline
Systems            & LMGP & LMGP0 & LWT  &\tfvw\   &WT	 \\ 
\hline 						
Mg$_8$	           & 0.18 & 0.63  & 1.19 & 1.09    &8.79 \\ 
Si$_8$	           & 0.22 & 2.17  & 4.86 & 1.46    &41.7 \\ 
Ga$_4$As$_4$       & 0.34 & 2.21  & 6.15 & 1.55    &51.8 \\ 
Mg$_{50}$          & 0.05 & 0.35  & 0.84 & 0.95    &3.23 \\ 
Si$_{50}$          & 0.11 & 0.95  & 3.73 & 1.53    &16.4 \\ 
Ga$_{25}$As$_{25}$ & 0.13 & 1.06  & 4.29 & 1.67    &22.7 \\ 
Mg$_8^{S}$            & 0.28 & 1.16  & 2.66 & 0.27    &19.4 \\ 
Mg$_8^{VS}$           & 0.09 & 1.67  & 3.88 & 0.10    &24.0 \\ 
\hline 						
\end{tabular}}
\end{table}

Table \ref{tab:Energy} shows that LWT considerably improves over WT. This indicates that the LDA procedure improves significantly the corresponding nonlocal KEDF with density independent kernel while at the same time removing the $\rho_0$ dependence in the functional. Interestingly, LMGP0 performs even better than LWT. This is an indication that the hypercorrelation term in the kernel further improves the performance of the functional. Adding the kinetic electron (i.e., the additional term $\omega_e(q)$ in the kernel, see \eqn{eq:kernel}), the LMGP functional achieves additional and important improvement over LMGP0, lining itself up to the KS-DFT results in an almost quantitative fashion. Strikingly, this is so despite the relatively uncomplicated formalism for the kinetic electron term in \eqn{eq:KE}.   

\begin{table}[!h]
\caption{\label{tab:Density} Mean unsigned relative error (MURE) for the electron density (measured by this quantity,  $\frac{1}{2N_{e}}\int |\rho_\text{OF-DFT}(\br)-\rho_\text{KS-DFT}(\br)|d\br$) in percentage points. Superscripts $S$ and $VS$ stand for ``strained'' and ``very strained'', respectively.}{\fontsize{10}{10}\selectfont 
\begin{tabular}{lccccc}
\hline \hline
Systems            & LMGP & LMGP0 & LWT  &\tfvw\  &WT	 \\ 
\hline 						
Mg$_8$	           & 3.79 & 4.12  & 4.05 & 11.36   &16.0 \\ 
Si$_8$	           & 4.84 & 4.90  & 4.74 & 8.28    &17.5 \\ 
Ga$_4$As$_4$       & 5.40 & 5.43  & 4.89 & 8.94    &19.3 \\ 
Mg$_{50}$          & 3.31 & 3.42  & 2.38 & 9.56    &10.3 \\ 
Si$_{50}$          & 4.59 & 4.65  & 3.60 & 7.24    &14.2 \\ 
Ga$_{25}$As$_{25}$ & 5.21 & 5.26  & 3.19 & 7.79    &16.8 \\ 
Mg$_8^{S}$            & 5.20 & 5.34  & 5.29 & 7.63    &18.6 \\ 
Mg$_8^{VS}$           & 3.94 & 4.10  & 4.87 & 5.60    &17.5 \\ 
\hline 						
\end{tabular}}
\end{table}

Reproducing the electron density is as important as reproducing the total energy. This was pointed out recently for exchange-correlation functionals \cite{Medvedev_2017} and it is even more important for KEDFs. Thus, we define $\frac{1}{2N_{e}}\int |\rho_\text{OF-DFT}(\br)-\rho_\text{KS-DFT}(\br)|d\br$, a measure of the electron density difference between KS-DFT and OF-DFT. The performance of the various functionals in reproducing the KS-DFT electron density are listed in Table \ref{tab:Density}. Once again, the three new functionals constitute an improvement over \tfvw\ and WT functionals.
We point out that although the total energies computed by \tfvw\ only partially differ from LWT, the LWT electron density is of much higher quality than the one from \tfvw.

In conclusion, we have addressed a long standing problem in the field of OF-DFT, i.e., the simulation of finite systems, such as quantum dots, reproducing benchmark KS-DFT results with unprecedented accuracy. This constitutes a major step forward for OF-DFT, a framework that was thought to only be applicable to bulk metals and alloys. Quantum dot structure prediction is now feasible with OF-DFT, opening the door to a new regime of applicability for this method. 

Our results are achieved by (1) imposing correct asymptotic behavior of the kinetic energy potentials, (2) accounting for nonlocality in the functional by construction, and (3) allowing the functional to adapt to systems with highly inhomogeneous electron density, via a technique inspired by the LDA approximation. Such a comprehensive, yet simple prescription leads to a numerically stable family of noninteracting kinetic energy functionals which we apply to 8 different quantum dots each realized in 100 different structures spanning energy windows ranging between 10 and 80 eV. Our most refined functional, LMGP, consistently reproduces the KS-DFT electronic energy for all 50-atom quantum dots to within 130 meV/atom. The energies of the 8-atom clusters are within 340 meV/atom of the KS-DFT reference. These errors are for the most part systematic in nature, as the OF-DFT energy values correlate almost perfectly to the KS-DFT benchmarks.

Although the noninteracting kinetic energy functionals presented here predict with unprecedented accuracy the total energy and electron density of the considered quantum dots, there is still room for improvement both in terms of computational accuracy as well as efficiency. Particularly, LMGP shows a significant improvement in comparison to LWT in terms of total energies, but struggles to improve the quality of the electron density. This indicates that the simple LDA approximation for the kernels can be further improved, for instance by including a dependency over the density gradient. This is currently being investigated.   

This material is based upon work supported by the National Science Foundation under Grant No. CHE-1553993.

\bibliography{./prg_bibliography/prg}
\end{document}